\documentclass[12pt]{article}
\usepackage{fleqn}
\usepackage{emlines2,bezier}
\usepackage{amssymb}
\author{S.S.Kokarev\\
Yaroslavl State Pedagogical University,\thanks{Russia, 150000,
Yaroslavl, Respublikanskaya 108, r.409}\\
Department of Theoretical and Experimental Physics}
\date{}
\title{SPACE-TIME AS STRONGLY BENT PLATE}
\DeclareSymbolFont{AMSa}{U}{msa}{m}{n}
\DeclareMathSymbol\square{\mathord}{AMSa}{"03} \global\let\Box\square

\begin{document}
\maketitle

\begin{abstract}
Futher development is made of a consept of space-time as
multidimensional elastic plate, proposed earlier in \cite{kok1,kok11}.
General equilibrium equations, including 4-dimensional tangent
stress tensor --- energy-momentum tensor of matter --- are derived.
Comparative analysis  of multidimensional elasticity theory (MET)
and GR is given. Variational principle, boundary conditions,
energy-momentum tensor, matter and space-time signature are reviewed
within the context of MET.
\end{abstract}

\section{Introduction}\label{intro}

It seems likely Cliffords works \cite{clif} laid philosophic foundations
of modern space-time physics. His ideas about
physical properties of space
have been realized in GR and in
a number of more recent works on its alternative formulations  and
generalizations. Part of this works has used analogy between GR and
{\it elasticity theory}, noted by Born \cite{born}. The
formal analogy between metric and strain tensor
and the idea  of elasticity of space-time have been discussed
in \cite{sah}-\cite{tart}.

The present paper joines the idea of elasticity with
a consept of {\it multidimensional space-time.}
This last is used in  majority of modern unified
theories (Kaluza-Klein models \cite{kal,wes,kok},
strings and $p$-branes
physics \cite{green}, supersymmetry \cite{super},
quantum field theory in bundle space \cite{bundle} and
others) in one way or another.
The matter of present paper corelates with multidimensional
theories, appealing to an idea of {\it embedding space.}
Embedding theory \cite{eiz} has been employed by many authors
in attempts of physical meaning of GR clearing \cite{jos}-\cite{pav}.

In works \cite{kok1,kok2} it has been proposed
to consider space-time as multidimensional elastic body,
whose sizes along four dimensions much greater,
then along another ones, so that this last are macroscopically
unobservable. Such body has been called {\it 4-dimensional plate.}
State of the plate can be described by multidimensional free elastic
energy, depending on mechanical strain of the plate.
Inner geometry, induced by the strain appears as geometry
on the plates surface, which can be easily obtained by standard
methods of embedding theory \cite{eiz}. In work \cite{kok1}
the case of a weak bending has been investigated.
Equilibrium equations, derived by variational method
have been disolved in \cite{kok2} for the case of
plane symmetry.

Preliminary comparing of the approach with standard GR implies
the hypothesis, which is  supported
by  following consideration: {\it gravity,
generating curvature of space-time,
is manifestation of normal bending of the plate.
Its energy is concentrated in microscopic
thicknesses of the plate.
Matter and fields appear as tangent to
the plate surface stresses and their energy
is macroscopic energy of stretches
and shears of the 4-dimensional cover.}
In other words {\it there is no space-time and matter separately,
but we have unified space-time-matter object}\footnote{The similar idea
has been proposed within Kaluza-Klein theory in \cite{wes,kok} for
different reasons and by quite different way.}.
The approach reveals ideas,
proposed by many other authors \cite{clif}-\cite{tart} and
\cite{jos}-\cite{pav}.

The plan of this paper is as follows.
First part is devoted to a generalization
of previous results, obtained in \cite{kok1}.

General mathematical and physical setting of the problem is given
in Sec.\ref{task}.

In Sec.\ref{prev} previous results of \cite{kok1}
are explicitly compared with linearized
Einstein theory.

In Sec.\ref{geneq} general equilibrium equations are derived
and in  Sec.\ref{clear}  meaning of strain weakness
conditions is cleared out.

Second part of the paper containes review of well known
objects of standard field theory in terms of multidimensional
elasticity theory (MET).

In Sec.\ref{lag} mechanical interpreting of field Lagrangian
is proposed. Example of classical newtral scalar field is
considered.

In Sec.\ref{cons} conservation laws and its origin are considered
from the view point of MET.

Sec.\ref{varpr} is devoted to a question of true variational
variables.

In Sec.\ref{an} notions of energy-momentum tensor, matter and its
density are analysed within the context of MET.
New understanding
of signature of space-time is given.

Sec.\ref{bd} is devoted to the problems of boundary terms and  boundary
conditions.

In Conclusion we make short discussion.

\section{Setting of the problem}\label{task}

Lets consider pseudoeuclidian vector space $M_{p+1,q+3}$
of $N+4$ dimensions\footnote{
Here and below we let big English letters represent $\overline{1,N+4}$,
small --- $\overline{1,N}$ and Greek
--- $0,1,2,3$, unless otherwise specified.},
$p+1$ of which are time-like
and $q+3$ --- space-like, $p+q=N$.
Span of four arbitrary linear independent vectors $\{{ t}_{(\mu)}\}$,
satisfying
the conditions\footnote%
{In present article for compactness and generality sakes
we use coordinateless
representation of tensors and tensor equations.
All tensor should be understood as
polylinear functionals, acting in corresponding vector and form spaces.
Because of metrics presence
we don't difer  vectors and forms by special notations,
unless otherwise specified.}:
\begin{equation}\label{sign}
{\Theta}({ t}_{(0)},{ t}_{(0)})>0,\ {\Theta}({
t}_{(i)},{ t}_{(i)})<0,\
\ (i=1,2,3),
\end{equation}
where
$\Theta$ --- metric in $M_{p+1,q+3}$,
forms
4-dimensional Minkowski plane $M_{1,3}$ and gives global
decomposition\footnote{We postulate,
that topology of $M_{p+1,q+3}$ is trivial.} of $M_{p+1,q+3}$:
$M_{p+1,q+3}=M_{1,3}\oplus M_{p,q}$, where $M_{p,q}$ ---
orthogonal ajunct of $M_{1,3}$ to original $M_{p+1,q+3}$.
Corresponding decomposition for metric has the form:
${\Theta}={\theta}+\bar{\theta}$,
where $\theta$ --- metric in $M_{1,3}$ and $\bar{\theta}$ --- metric
in $M_{p,q}$, so that $\bar\theta(u, )=0$ and $\theta(v,)=0$ for any
vectors $u\in M_{1,3},\ v\in M_{p,q}$.

Let $\{X^A\}$ be the set of affine coordinates\footnote{
The $\{X^A\}$  forms vector $X$ with respect to affine group
of transformations in $M_{p+1,q+3}$.}
in $M_{p+1,q+3}$,
concordant with the decomposition of $M_{p+1,q+3}$:
$X=x\oplus\bar x$, where
$\{x^{\mu}\}$ and $\{\bar x^m\}$ ---  the sets of coordinates
in $M_{1,3}$ and $M_{p,q}$ correspondingly, satisfying the condition
\begin{equation}\label{ort}
\Theta(dx,d\bar x)=0;\ \ \theta(d\bar x,d\bar x)=0;\ \
\bar\theta(dx,dx)=0,
\end{equation}
Then equation of $M_{1,3}$ will take the simple form:
\begin{equation}\label{pleq}
\bar{ x}=0.
\end{equation}
and line element
in $M_{p+1,q+3}$
due to conditions (\ref{ort}),
takes the form:
\begin{equation}\label{line}
ds^2={\Theta}(dX,dX)={\Theta}(dx+d\bar x,dx+d\bar x)
={\theta}(d x,d x)+\bar{\theta}(d\bar x,d\bar x).
\end{equation}
Without loss of generality
metrics ${\theta}$ and $\bar{\theta}$ can be put to diagonal form
with $\pm1$ on diagonals:
$$
\|\theta\|={\rm diag}(1,-1,-1,-1);\ \ \
\|\bar\theta\|={\rm diag}(\varepsilon_1,\dots,\varepsilon_N);\ \
\varepsilon_i=\pm1
$$
by independent linear transformations of $\{x^{\mu}\}$ and $\{\bar x^m\}$
and  these last may be regarded as cartesian.

We postulate, that {\it geometrical Minkowski plane $M_{1,3}$
is middle plane of 4-dimensional physical elastic plate},
which models classical macroscopic Minkowski space-time of special relativity.
So, we endow $M_{1,3}$ by the set of thicknesses $\{h_m\}$
in extradimensions, lied in $M_{p,q}$, and by multidimensional
phenomenological elastic constants --- Lame coefficients (see
(\ref{fren})).

Let consider smooth\footnote{Requirement $\Xi\in C^4$ is sufficient.}
vector field $ \Xi$,
defined on $M_{1,3}$: ${\Xi}={ \Xi}({ x})$. This field,
if it is nonlinear function of $ x$, gives transformation
of $M_{1,3}$: $M_{1,3}\stackrel{\Xi}{\rightarrow}V_{1,3}$,
where $V_{1,3}$ --- riemannian space, which
{\it models macroscopic space-time of GR.}  Such transformation is
deformation, which from  physical point of view
can be associated with mechanical {\it straining} of
the plate\footnote{Here we don't pose the question about nature
of multidimensional straining forces.}.
So $\Xi$ can be treated as displacement vector field of standard
elasticity theory \cite{land1}, $V_{1,3}$ --- as middle plane
of strained plate and $M_{1,3}$  --- as coordinate
map of $V_{1,3}$.
Position of any point of $V_{1,3}$ in $M_{p+1,q+3}$ after straining
is described by the formula:
$$
X'=X+\Xi
$$
or, using (\ref{pleq}), in splitting form
$$
x'=x+\xi;\ \ \bar x'=\bar\xi,
$$
where $\xi$ and $\bar\xi$ --- are tangent and normal to
$M_{1,3}$ components of $\Xi$.

Line element (\ref{line}) on the plate surface with using
(\ref{ort}) can be transformed to
\begin{equation}\label{metr}
ds'^2={\Theta}(d{ X}+d{ \Xi},d{
X}+d{\Xi})=ds^2+2{\cal D}(d{ x},d{ x})=g(dx,dx),
\end{equation}
where
\begin{equation}\label{strain}
{\cal D}={\partial}\dot\otimes{\xi}+
\frac{1}{2}(\partial'\otimes\partial'')(
\theta(\xi',\xi'')+
\bar\theta(\bar\xi',\bar\xi''))
\end{equation}
--- is symmetric {\it strain tensor},
$g=\theta+2{\cal D}$ --- metric on $V_{1,3}$.
 We have used following notations:
$\partial$ --- partial derivative operator in $M_{1,3}$,
${ a}\dot\otimes { b}\equiv(1/2)({ a}\otimes{ b}+{
b}\otimes{ a})$ --- symmetric tensor
product, $'$ and $''$ denote arguments of partial
derivatives.
In coordinate form (\ref{strain}) is:
\begin{equation}\label{strainc}
{\cal
D}_{\mu\nu}=\xi_{(\mu,\nu)}+\frac{1}{2}\xi^{\lambda}_{,\mu}\xi_{\lambda,\nu}+
\frac{1}{2}\bar\xi^m_{,\mu}\bar \xi_{m,\nu},
\end{equation}
where Greek indices are  contracted with the help of metric ${\theta}$,
small latin indices --- with the help of  $\bar{\theta}$.

We have formulated kynematic part of the problem.
At the first step of dynamic part we postulate
multidimensional {\it Hookes law}.
In other words,
we take free elastic energy density in the form:
\begin{equation}\label{fren}
{\cal F}=\mu{\cal D}\cdot{\cal D}+\frac{\lambda}{2}\theta^2({\cal D})
\end{equation}
where
$\mu,\lambda$ --- phenomenological multidimensional
{\it Lame coefficients},
${\cal D}\cdot{\cal D}\equiv {\cal D}_{\mu\nu}{\cal D}^{\mu\nu}$
--- tensor "scalar product", $\theta({\cal D})\equiv{\cal
D}_{\mu}^{\mu}$ --- trace of ${\cal D}$.
In standard elasticity theory the constants $E$ --- {\it Youngs
modulus} and $\sigma_P$ --- {\it Poisson coefficient}
are used more often, because of their physical clearness.
They are related with Lame coefficients by the formulas (see
\cite{kok1}, also \cite{tart}):
\begin{equation}\label{coeff}
\mu=\frac{E}{2(1+\sigma_P)};\ \
\lambda=\frac{E\sigma_P}{(1+\sigma_P)(1-(n-1)\sigma_P)};
\end{equation}
In our case\footnote{Under $n=3$ we get standard formulas of
\cite{land1}.}
$n=N+4$.

Direct generalization of standard thermodynamics gives
\begin{equation}\label{term}
d{\cal F}=-{\cal S}dT+\sigma\cdot d{\cal D}
\end{equation}
where ${\cal S}$ and $T$ --- multidimensional entropy density and temperature
of plate substance, $\sigma$ --- {\it stress tensor}.
Its alternative definition can be expressed by
the integral
\begin{equation}\label{stress}
\oint\limits_{\partial\Sigma}\sigma(\ ,d\,{\rm vol}[\partial\Sigma])=f_{\Sigma}
\end{equation}
where $f_{\Sigma}$  --- multidimensional force vector, acting on
finite arbitrary volume $\Sigma$ inside elastic body,
$d\,{\rm vol}[\partial\Sigma]$ ---  suitable element of boundary hyphersurface
$\partial\Sigma$, oriented out of $\Sigma$.
Expressions (\ref{term}) and (\ref{stress})
give equivalent $\sigma$ if last is symmetric,
that can be accepted without loss of generality (see \cite{land1}).
Divergence of $\sigma$ determines local volume force density $f$:
\begin{equation}\label{densf}
{\rm div}\, \sigma=f,
\end{equation}
where ${\rm div}\, \sigma_{\alpha}=\partial^\beta\sigma_{\alpha\beta}$.
As it is in standard theory expression
(\ref{term}) gives the rule
for  calculation of stress tensor
\begin{equation}\label{calc}
\sigma=\left(\frac{\partial{\cal F}}{\partial{\cal D}}\right)_{T}.
\end{equation}
Taking differential of (\ref{fren})
$$
d{\cal F}=2\mu{\cal D}\cdot d{\cal D}+\lambda\theta({\cal
D})d\theta({\cal D})
$$
and substituting it into (\ref{calc}) we obtaine
\begin{equation}\label{hook}
\sigma=2\mu{\cal D}+\lambda\theta({\cal D})\theta
\end{equation}
--- {\it multidimensional Hookes law}.

If $\Pi$ is vector force density, acting at the
bound $\partial\Sigma$ of multidimensional elastic body, then boundary
conditions for $\sigma$ are:
\begin{equation}\label{bound}
\sigma(\ ,\tau)|_{\partial\Sigma}=\bar P;\
\ \sigma(\ ,n)|_{\partial\Sigma}=P,
\end{equation}
where $n,\tau,P,\bar P$ --- normal and tangent
to $\partial\Sigma$ unit vectors and components of $\Pi$
correspondingly.
As in standard theory we assume, that external bending
stresses are much less then internal compensating ones.
In other words we put:
\begin{equation}\label{bound1}
\sigma(\ ,\tau)|_{\partial\Sigma}=\sigma(\ ,n)|_{\partial\Sigma}=0.
\end{equation}

Note, that all differentiations and integrations can
be expressed through internal coordinates
$x'$
of the plate surface
by changing
$\partial\to\nabla$ and by definition of suitable
volume form, for example: $d\,{\rm vol}[\Sigma]=\sqrt{{\rm det}[g]}
dx'^0\wedge dx'^1\wedge dx'^2\wedge dx'^3.$ Fortunately this is
not necessary, since all corrections
are of higher smallness and can be ommited in linear elasticity
theory.

All relations (\ref{strain})-(\ref{bound1}), being general in its form,
have, as a matter of fact, 4-D sense. Particulary,
(\ref{fren}) expressed through (\ref{strainc}) gives only 4-dimensional
part of free energy, induced by 4-dimensional stretches and
shears of the plate, which yet can be regarded
as elastic 4-dimensional cover without thicknesses.
By these reasons  we shall use notation ${\cal F}_s$
instead of ${\cal F}$ to difer it from ${\cal F}_b$
--- energy density of {\it pure bending}.

\section{Previous results}\label{prev}

To take into account free energy of pure bending it is necessary to
investigate stress distribution along extradimensions.
It has been carried out in \cite{kok1}.
We present final results without intermediate calculations.

Free energy of bending, calculated with using
(\ref{bound1}) and integrated over thicknesses
of the plate has the form
\begin{equation}\label{frenb}
F_{b}=\frac{\mu H_{N}}{12}\int\limits_{\Sigma}
\left\{\bar\theta(\partial^2\bar\xi_h{\cdot}\partial^2\bar\xi_h)
+f\bar\theta(\theta(\partial^2\bar\xi_h),\theta(\partial^2\bar\xi_h))\right\}
d\,{\rm vol}[\Sigma],
\end{equation}
where  $H_{N}=\prod\limits_{i=1}^{N}h_{i}$, factor
$f=\lambda/(N\lambda+2\mu)$,
$\partial^2$ --- tensor second derivative operator:
$[\partial^2f(x)]_{\alpha\beta}\equiv\partial_{\alpha}\partial_{\beta}f(x)$,
$\bar\xi_h=\{\bar\xi^m\}\times\{h_m\}\equiv\{\bar\xi^m h_m\}$ (without
summation), $\Sigma$ --- 4-dimensional region of the plate.
Note, that (\ref{frenb}) does'nt contain $\xi$-component,
which can be ommited under weak bending (see \cite{land1}).

Variation of (\ref{frenb}) over $\bar\xi$ (under condition
$\delta(d\, {\rm vol}[\Sigma])\approx0$)
is
\begin{equation}\label{var}
\delta F_{b}=
\int\limits_{\Sigma}\bar\theta(\Box^{2}\bar\xi_{D}-\bar
P,\delta\bar\xi)\, d\,{\rm vol}[\Sigma]-
\oint\limits_{\partial\Sigma}\bar\theta(\theta(\Box\partial\bar\xi_{D},
d\,{\rm vol}[\partial\Sigma]),\delta\bar\xi)
\end{equation}
\[
+\frac{1}{f+1}
\oint\limits_{\partial\Sigma}
[f\bar\theta(\Box\bar\xi_{D},\theta(d\,{\rm vol}[\partial\Sigma],\delta\partial\bar\xi))
+\bar\theta(\partial^2\bar\xi_D(d\,{\rm vol}[\partial\Sigma],\delta\partial),\bar\xi)]
\]
where
$\Box=\theta(\partial,\partial)$ --- wave operator,
$\bar\xi_D=\{\bar\xi^m\}\times\{D_m\}=\{\bar\xi^m D_m\}$ (without
summation),
\begin{equation}\label{stiff}
D_{m}=\frac{\mu
H_{N}h_{m}^{2}(f+1)}{6}=\frac{EH_{N}h_{m}^{2}}{12(1+\sigma_P)}
\frac{1+\sigma_P(N-n+2)}{1+\sigma_P(N-n+1)}
\end{equation}
--- constant cylindrical stiffness factor of the plate in $m$-th
extradimension. In our case $n=N+4$\footnote{For $n=3$ $N=1$
we have the standard formulae \cite{land1}}.
$\bar P$ --- external bending force density, satisfying
$\theta(\bar P,\ )=0$
and
$$
\delta U=-\oint\limits_{\Sigma}\bar\theta(\bar P,\delta\bar\xi)d\,{\rm vol[\Sigma]},
$$
where $U$ --- multidimensional potential energy.
Last two integrals in (\ref{var}) are
boundary terms, which must be taken into account
to obtaine unique solution.
Equilibrium equations, derived from
extremum condition $\delta(F_b+U)=0$ are inhomogeneous
bewave equations
\begin{equation}\label{bewave}
\Box^2\bar\xi_D=\bar P
\end{equation}
--- direct generalization of Sophy-Germain
beharmonic equation: $D\Delta^2\xi=P$ of standard elasticity
theory \cite{land1}.

Using dimensional arguments, one can get the following
remarkable relation between multidimensional parameters
and Einstein gravitational constant $\ae$:
$$
Eh^{N+3}\sim1/\ae.
$$
This relation is in line with Sacharov's hypothesis
of possible elasticity of space-time \cite{sah}.
Another variants of possible relations see in \cite{kok1}.

In case $f=-1$ $(\sigma_P=1/2)$ integrand bracket in (\ref{frenb})
become similar to ${\cal R}_s$ --- scalar curvature of $V_{1,3}$,
expressed through second derivatives $\partial^2\bar\xi$
with using embedding theory  \cite{eiz}. In \cite{kok1} this fact has been
interpreted as correspondence between the proposed approach and
linearized Einstein theory. It can be checked
by the straightfoward way.

Away from a bodies,
gravitational field can be described by linearized theory
of gravitation, wherein metric tensor is the sum of a flat
Minkowski metric and its small disturbance $\psi$ \cite{land2}:
\begin{equation}\label{linear}
g=\theta+\psi.
\end{equation}
In this case
one may reject qudratic over Cristoffel symbols part
of expression for curvature tensor ${}^4{\cal R}$
and hold only linear over second derivatives $\partial^2g$
one, that leads to the expression:
\begin{equation}\label{lcurv}
{}^4{\cal R}^{\rm lin}(a,b,c,d)=
\frac{1}{2}(\partial^2\psi(a,d,b,c)+\partial^2\psi(b,c,a,d)-
\partial^2\psi(a,c,b,d)-\partial^2\psi(b,d,a,c)),
\end{equation}
where
$\partial^2\psi(a,b,c,d)\equiv\partial_{\alpha}\partial_{\beta}\psi_{\gamma\delta}
a^{\gamma}b^{\delta}c^{\alpha}d^{\beta}$ and $a,b,c,d$ --- arbitrary
vectors.
Then, using expression (\ref{metr})
for riemannian metric of strained plate
we get $\psi$ in the form:
\begin{equation}\label{metrl}
\psi=\frac{1}{2}\bar\theta(\partial\bar\xi,\partial\bar\xi)
\end{equation}
where  we have ommited
addends with $\xi$
in (\ref{metr}).

Substituting this expression into (\ref{lcurv})
and making elementary transformations we have
\begin{equation}\label{rieml}
{}^4{\cal R}^{\rm lin}(a,b,c,d)=
(\bar\theta(\partial^2\bar\xi(a,c),\partial^2\bar\xi(b,d))-\bar\theta(\partial^2
\bar\xi(a,d),\partial^2\bar\xi(b,c)),
\end{equation}
that after appropriate contractions with flat metric $\theta$
(and rescaling $\bar\xi\to\bar\xi_h$) gives (up ty a sign)
the integrand bracket of
(\ref{frenb})
under $f=-1$.

So, multidimensional Hookes law, applied for describing
of weakly bent 4-dimensional plate under special value
of elastic constants corresponds to linearized
GR.

\section{General equilibrium equations}\label{geneq}

To consider non weak gravitational and another 4-dimensional fields
one should take into account both free energy $F_b$ of pure bending
and free energy $F_s$ of 4-dimensional stretches-shears.
Our derivation of general equilibrium equation follows to \cite{land1}
with suitable generalizations and addends.

Total elastic free energy $F$ is the sum:
$F=F_{s}+F_{b}$, where second term is (\ref{frenb}).
Stretch energy $F_s$ can be obtained from (\ref{fren})
by integrating over plate volume. Using change
$d\,{\rm vol}[M_{p+1,q+3}]\to H_Nd\,{\rm vol}[\Sigma]$ we get
\begin{equation}\label{ifren}
F_s=H_N\int\limits_{\Sigma}{\cal F}_s\,d\,{\rm vol}[\Sigma].
\end{equation}
Variation of $F_b$ over $\bar\xi$ is (\ref{var}).
Variation $F_s$ depending on both $\xi$ and $\bar\xi$
is
\begin{equation}\label{vars}
\delta F_s=H_N\int\limits_{\Sigma}\delta{\cal F}_s\,d\,{\rm vol}[\Sigma]=
H_N\int\limits_{\Sigma}\frac{\partial{\cal F}_s}{\partial{\cal D}}\cdot
\delta{\cal D}d\,{\rm vol}[\Sigma]=H_N\int\limits_{\Sigma}\sigma\cdot\delta{\cal
D}d\,{\rm vol}[\Sigma],
\end{equation}
where (\ref{fren}) and (\ref{calc}) have been used.
Substituting (\ref{strain}) (ommiting second term (see
Sec.\ref{clear}))
into (\ref{vars})
and carring out standard integration by parts
we get
\begin{equation}\label{varsf}
\delta F_s
=-H_N\int\limits_{\Sigma}[\theta({\rm div}\,\sigma,\delta\xi)+
\bar\theta(\theta(\partial,\sigma(\, ,\partial\bar\xi)),\delta\bar\xi)]d\,{\rm vol}
[\Sigma]+
\end{equation}
\[
H_N\oint\limits_{\partial\Sigma}[\sigma(d\,{\rm vol}[\partial\Sigma],\delta\xi)+
\sigma(\bar\theta(\partial\bar\xi,\delta\bar\xi),d\,{\rm vol}[\partial\Sigma])],
\]
where last integral is boundary term.

If external  force has potential $U$,
then
\begin{equation}\label{varpot}
\delta U=-\int\limits_{\Sigma}\bar\theta(\bar P,\delta\bar\xi)d\,{\rm vol}
[\Sigma]-
\int\limits_{\Sigma}\theta(P,\delta\xi)d\,{\rm vol}[\Sigma],
\end{equation}
where $P+\bar P=\Pi$ --- external force density, $P$ --- stretching
and $\bar P$ --- bending parts.
Finally, extremality condition $\delta(F+U)=0$
gives equilibrium equations
\begin{equation}\label{perpeq}
\Box^2\bar\xi_D-H_N\theta(\partial,\sigma(\ ,\partial\bar\xi))=\bar P
\end{equation}
\begin{equation}\label{pareq}
H_N{\rm div}\,\sigma=-P
\end{equation}
or in coordinate form:
$$
\Box^2\bar\xi_D^m-H_N(\sigma^{\alpha\beta}\bar\xi^m_{,\beta})_{,\alpha}
=\bar P^m;\ \
H_N\sigma^{\alpha\beta}_{,\beta}=-P^{\alpha},
$$
and boundary terms
\begin{equation}\label{bgen}
\oint\limits_{\partial\Sigma}\bar\theta(H_N\sigma(\partial\bar\xi,d\,{\rm
vol}[\partial\Sigma])-\theta(\Box\partial\bar\xi_D,d\,{\rm
vol}[\partial\Sigma]),\delta\bar\xi)+
\end{equation}
\[
\frac{1}{f+1}\oint\limits_{\partial\Sigma}[f\bar\theta(\Box\bar\xi_D,
\theta(d\,{\rm vol}\partial\Sigma),\delta\partial\xi))+
\bar\theta(\partial^2\bar\xi_D(d\, {\rm
vol}[\partial\Sigma],\delta\partial),\bar\xi)]+
\]
\[
H_N\oint\limits_{\partial\Sigma}\sigma(d\,{\rm
vol}[\partial\Sigma],\delta\xi).
\]
The obtained equations (\ref{perpeq})-(\ref{pareq})
describe selfconsistent system of tangent and orthogonal
to $M_{1,3}$  straines. Eq.(\ref{perpeq})
plays the similar to Einstein equations role: under
$\bar P=0$ induced by $\bar\xi$ geometry
is "determined" by the physics, concluded in $\sigma$.
Eq.(\ref{pareq}), in turn, describes 4-dimensional physics,
formed by tangent to the plate surface stresses.
If one take into account identity of stress tensor $\sigma$ and
energy-momentum tensor $T$ of GR, then (\ref{pareq})
can be interpreted as equation of motion of matter in
$V_4$. When $\sigma=0$ we obtain approximative
bewave equations (\ref{bewave}) of weak bending case.

\section{Weakness conditions}\label{clear}

Lets clarify meaning of straines weakness condition,
which provides validity of Hookes law. One should difer two
aspects of the condition.

1. Smallness of  strain tensor components lets to use
simple quadratic expression (\ref{fren}) for free energy density,
which can be interpreted as first member of Taylor decomposition.
This condition is expressed by the inequalities with partial
derivatives:
$$
|\partial\xi|\ll1;\ \ |\partial\bar\xi|\ll1.
$$
which means smallness of lengths and angles variation
under straining $\Xi$\footnote{Note, that smallness
of $\partial\bar\xi$ is not so necessary. For example,
$\bar\xi=\theta(A,x)$ with arbitrary constant vector $A$
determines rigid rotation of $M_{1,3}$ in embedding space.}.
Since $\xi$ and $\bar\xi$
are independent, then there is two independent smallness
parameters: $|\partial\xi|\sim\varepsilon$ and
$|\partial\bar\xi|\sim\bar\varepsilon$.
So, algebraic structure of strain tensor (\ref{strain})
is
$$
{\cal D}\sim\varepsilon+\varepsilon^2+\bar\varepsilon^2.
$$
Easily to see, that second term is of higher smallness
in comparison with the first under any relation between $\varepsilon$
and $\bar\varepsilon$.
Denoting $\varepsilon+\bar\varepsilon^2=\delta_1$ we get for the metric
(\ref{metr}):
$$
g\sim1+\delta_1.
$$
So, in linear approximation all contractions can be made
with the help of flat metric
$\theta$ and $\delta(d\,{\rm vol}[\Sigma])$ can be taken
zero.

\vspace{0.5cm}

\begin{figure}[htb]
\begin{center}
\unitlength=1.00mm
\special{em:linewidth 0.4pt}
\linethickness{0.4pt}
\begin{picture}(30.33,48.33)
\emline{5.00}{42.00}{1}{18.00}{3.00}{2}
\emline{18.00}{3.00}{3}{30.00}{42.00}{4}
\bezier{112}(4.67,42.33)(17.33,48.33)(30.33,42.33)
\bezier{76}(9.00,30.33)(19.00,33.67)(26.33,30.33)
\emline{11.67}{37.00}{5}{15.67}{37.67}{6}
\emline{16.67}{37.67}{7}{19.33}{37.67}{8}
\emline{20.67}{37.33}{9}{24.33}{36.67}{10}
\put(17.00,40.33){\makebox(0,0)[cc]{$\delta l_0$}}
\put(16.67,48.00){\makebox(0,0)[cc]{$\delta l_{+}$}}
\put(17.33,29.00){\makebox(0,0)[cc]{$\delta l_{-}$}}
\put(29.67,34.67){\makebox(0,0)[cc]{$h$}}
\put(25.67,20.67){\makebox(0,0)[cc]{$R$}}
\emline{6.33}{38.67}{11}{10.00}{44.33}{12}
\emline{8.00}{34.00}{13}{15.33}{45.33}{14}
\emline{9.67}{30.67}{15}{15.00}{39.00}{16}
\emline{18.33}{43.33}{17}{19.33}{45.33}{18}
\emline{15.00}{32.00}{19}{23.00}{44.67}{20}
\emline{19.00}{32.00}{21}{26.00}{43.67}{22}
\emline{22.67}{31.67}{23}{29.33}{42.67}{24}
\emline{25.67}{31.00}{25}{27.33}{33.67}{26}
\emline{7.33}{35.67}{27}{10.33}{36.33}{28}
\emline{26.33}{36.00}{29}{28.00}{35.33}{30}
\end{picture}
\end{center}
\caption{Infinitesemal part of a bent plate.}\label{bending}
\end{figure}
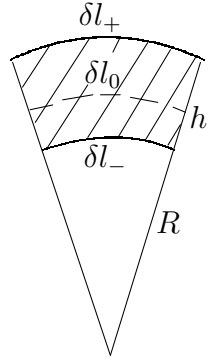

2. Second condition --- bending weakness.
Fig.\ref{bending}
shows 2-dimensional section of bent plate with magnituded
thickness; $\delta l_0$ is invariable line element, lying
at newtral surface, $\delta l_{\pm}$ ---
line elements, lying at opposite sides of the plate and
obtained by positive and negative srtetches of $\delta l_0$.

Hookes law validity means, that relative length variations along thickness $h$
\begin{equation}\label{small}
\frac{\delta l_{+}-\delta l_{-}}{\delta l_{+}+\delta l_{-}}\ll1.
\end{equation}
It is easily to find, that
$$
\frac{\delta l_{+}}{\delta l_{-}}=\frac{R+(h/2)}{R-(h/2)},
$$
where $R$ --- curvature radius of newtral surface in embedding
space.
After substituting into (\ref{small}),
we get: $h/R\ll1$. Using embedding  theory relation
$1/R^2\sim(\partial^2\bar\xi)^2$,
we get for multidimensional plate
\begin{equation}\label{smbend}
h^2(\partial^2\bar\xi)^2=\delta^2_2\ll1.
\end{equation}
In other words, we have the set $\{\delta_2^{m}\}$ of smallness
parameters --- each determines bending smallness in corresponding
extradimension.

Now we can compare contributions of $F_b$ and $F_s$ in total
expression $F$.
Namely,
${\cal D}\sim\delta_1$, $\sigma\sim E\delta_1$,
so for integral (\ref{ifren}) we get estimation:
$$
F_s\sim E H_NV\delta_1^2,
$$
where $V$ --- 4-dimensional square (volume) of the plate.
For integral
(\ref{frenb}) we get
$$
F_b\sim\mu H_NV\sum\limits_{m=1}^Nh_m^2(\partial^2\bar\xi^m)^2\sim
EH_NV\sum_{m=1}^N(\delta_{2}^m)^2.
$$

So, the two components of total $F$ has independent smallness rates.
In case $(\delta_1/\delta_2^m)^2\gg1$ for all $m$ $F_b$
can be ommited and we have 4-dimensional physics in flat space-time.
In case $(\delta_1/\delta_2^m)^2\ll1$ for any $m$ $F_s$ can
be ommited and we have curved vacuum space-time.
When $(\delta_1/\delta_2^m)^2\sim1$ for some $m$
both the energies should be hold and we have curved space-time
with fields and matter. Intermediate situations are
possible, when different members of $\{\delta_2^m\}$
have different smallness rates.

\section{New treatment of Lagrangian formalism}\label{lag}

Obtained results lead to  curious interpretation of  well known
objects of standard field theory. Assume, that in
(\ref{strain}) $\bar \xi=0$ and we have 4-dimensional
physics in Minkowski space-time.

First of all, lets call attention on
analogy of the two expressions
\begin{equation}\label{iden}
\delta F=\int\sigma\cdot\delta{\cal D}\, d\,\mbox{\rm vol},\ \
\mbox{\rm and}\ \
\delta S=\frac{1}{2c}\int T\cdot\delta g\, d\,\mbox{\rm vol}.
\end{equation}
where $d\,\mbox{\rm vol}$ denotes suitable form of volume.
The first is general thermodynamic relation, connecting stress
tensor with infinitisemal variation of an elastic free energy
(see (\ref{term}) and (\ref{calc})).
The second --- is well known
rule for calculation of symmetric energy-momentum tensor
of fields or matter from its lagrangian density $\cal L$: $S=
\int{\cal L}\, d\,\mbox{\rm vol}$.

This analogy can be understood if
one take into account the two points: 1)
$\sigma$ and $T$ have the same physical meaning;
2) tensor $\sigma$ and consequtely $T$ are generated by tangent
straining of plate medium. If so, then free energy density
of stretches ${\cal F}_s({\cal D})$ and matter lagrangian
${\cal L}(q,\partial q)$, depending on dynamical field variables
$q\equiv\{q^m\}$ and its derivatives should be identified
(up to a dimensional constant): ${\cal F}=\kappa{\cal L}$.
Futhermore, variables $q$ acquire the following mechanical sense:
they detrmine the way  of description of $M_{1,3}$ tangent straining
or in mathematical form: ${\cal D}={\cal D}(q)$.

\begin{figure}[htb]
\begin{center}
\unitlength=1mm
\special{em:linewidth 0.4pt}
\linethickness{0.4pt}
\begin{picture}(128.67,21.67)
\emline{5.67}{5.67}{1}{24.00}{21.67}{2}
\emline{24.00}{21.67}{3}{57.00}{21.67}{4}
\emline{57.00}{21.67}{5}{41.00}{6.00}{6}
\emline{41.00}{6.00}{7}{5.33}{6.00}{8}
\emline{77.33}{5.33}{9}{95.67}{21.33}{10}
\emline{95.67}{21.33}{11}{128.67}{21.33}{12}
\emline{128.67}{21.33}{13}{112.67}{5.67}{14}
\emline{112.67}{5.67}{15}{77.00}{5.67}{16}
\emline{52.33}{13.00}{17}{83.00}{13.00}{18}
\emline{83.00}{13.00}{19}{79.00}{14.67}{20}
\emline{83.00}{13.00}{21}{79.00}{11.33}{22}
\put(66.33,15.67){\makebox(0,0)[cc]{active}}
\put(66.33,10.33){\makebox(0,0)[cc]{transformation}}
\emline{11.33}{10.67}{23}{45.33}{10.67}{24}
\emline{18.67}{17.00}{25}{52.00}{17.00}{26}
\emline{34.00}{21.67}{27}{19.00}{6.00}{28}
\emline{44.00}{21.67}{29}{31.00}{6.00}{30}
\bezier{200}(83.33,10.67)(89.67,7.33)(98.67,10.33)
\bezier{200}(99.33,10.67)(109.33,13.33)(117.00,10.67)
\bezier{200}(90.67,17.00)(92.67,14.33)(103.33,17.00)
\bezier{200}(104.00,17.33)(114.33,20.33)(124.00,17.00)
\bezier{200}(104.33,21.33)(97.33,18.00)(98.33,13.67)
\bezier{200}(98.67,13.33)(100.67,8.00)(89.67,6.00)
\bezier{200}(114.67,21.67)(105.00,17.00)(108.67,13.67)
\bezier{200}(109.00,13.33)(111.33,9.00)(102.67,6.00)
\put(23.67,3.00){\makebox(0,0)[cc]{(a)}}
\put(96.33,3.00){\makebox(0,0)[cc]{(b)}}
\end{picture}
\end{center}
\caption{Mechanical interpreting of fields and matter.}\label{matter}
\end{figure}
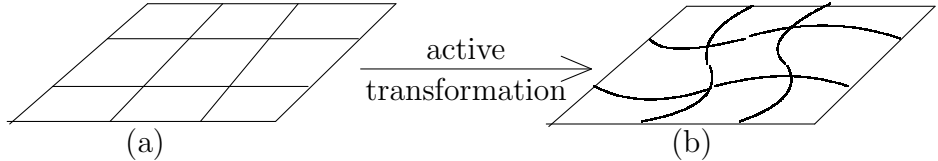
When matter is absent, we have unstrained Minkowski
plane  $M_{1,3}$  (Fig.\ref{matter}(a)) with standard Minkowski metric $\theta$.
After diffeomorphism $x\to x'=x'(x)$, which has
not passive (as in GR), but active sense of real flat straining
of the plate medium we obtain stressed plate $M'_{1,3}$ with curved
coordinate lines (Fig.\ref{matter}(b)).
In spite of its inner geometry is not change,
all metric relations should be represented in general covariant
form with using new metric $g=\theta+2{\cal D}$
in accordance with  (\ref{strain}).
Both $\cal F$ and $\cal L$ must implicitly containe metric $g$
(to get from $\cal D$ or $q,$ and $\partial q$ scalar
expressions).
Note, that if straining is not strong, then $g$ can be changed by $\theta$
in all contractions.
From the kind of $g$ we get $\delta g=2\delta{\cal D}$,
and so
\begin{equation}\label{iden1}
\sigma=\frac{\delta{\cal F}}{\delta{\cal D}}=2\frac{\delta{\cal
F}}{\delta g}\sim\frac{\delta{\cal L}}{\delta g}=T.
\end{equation}
If this analogy is not accident, then {\it any classical field
lagrangian can be treated as free elastic energy density
and specific choice of field variables
is determined by the kind of plate straining.}

Lets demonstrate it by example with scalar field.
Consider the following active transformation $M_{1,3}\to M'_{1,3}$:
\begin{equation}\label{tr}
\theta\to g=(1+2\varepsilon_1\phi)\theta, \ \
x'=x+\varepsilon_2a(\phi).
\end{equation}
where $\phi$ --- dilaton scalar field of small value,
$a$ --- vector field, depending on
coordinates only through $\phi$ and satisfying
the conditions $(1)\ a(0)=0,\ (2)\ \theta(a_{\phi},a_{\phi})=c=const\
(a_{\phi}\equiv da/d\phi)$ and
$(3)\ \mbox{\rm div}\, a=0$\footnote{Proof of existence of $a$
with conditions (1)-(3) for any $\phi$ is given in Appendix.}.
First transformation is conformal transformation of metric,
which determines homogeneous stretch of $M_{1,3}$,
and second --- associated with $\phi$ special (stretchless) translation.
Factors $\varepsilon_1,\ \varepsilon_2$ can be taken
$0$ or $1$ independently: they will allow us to distinguish role
of dilatation and translation in final expression.
Metric $g$ of $M'_{1,3}$ is
\begin{equation}\label{dil}
ds'^2=g(dx',dx')=(1+2\varepsilon_1\phi)\theta(dx+\varepsilon_2a_{\phi}
d\phi,dx+\varepsilon_2a_{\phi}d\phi)=ds^2+2{\cal D}(dx,dx),
\end{equation}
where strain tensor
$${\cal
D}=\varepsilon_1\phi\theta+\varepsilon_2a_{\phi}\stackrel{\cdot}{\otimes}\partial\phi,
$$
and quadratic over $\phi$ and $\partial\phi$ addends are
ommited.
Free energy density (\ref{fren}) with using
conditions (2) and (3) on $a$ takes the form
\begin{equation}\label{frensf}
{\cal F}=\frac{\varepsilon^2_2}{2}c\mu(\partial\phi)^2+4\varepsilon_1^2(\mu+2\lambda)\phi^2.
\end{equation}
Comparing with standard lagrangian of massive newtral scalar field
$$
{\cal L}_{\rm sf}=\frac{1}{2}(\partial\phi)^2-\frac{1}{2}m^2\phi^2,
$$
we get relation for mass\footnote{From (\ref{coeff}) it follows, that
$m=0$ if $\sigma=1/N$.} $m^2\sim\mu+2\lambda$.
When $\varepsilon_1=0$, we have only kynetic term.
In case $\varepsilon_2=0$ we get ${\cal F}\sim\phi^2$.

\section{Equilibrium equations and  conservation laws}\label{cons}

It is well known, that
extremality of $F[\Xi]$ and $S[q]$ leads
to Euler-Lagrange equations, which
in turn, provide validity of  equilibrium
equations in the first case and  conservation laws
in the second \cite{land2,bogol}:
\begin{equation}\label{conserv}
\delta F=0\ \to \
{\rm div}\, \sigma=0\ \ \
\delta S=0\ \to \ {\rm div}\, T=0;\
\end{equation}
In line of present investigation it would be naturally
to use unified language and regard conservation
laws  as equilibrium equation of some elastic body.
In view of results of Sec.\ref{geneq}
we can conclude, that  {\it this  body is space-time itself.}

When dynamical system is not closed,
we write ${\rm div}\,\sigma=f$, where $f$ --- vector of external force density.
If $f$ can be represented\footnote{This is possible,
when interaction is local.}
as $-{\rm div}\,\sigma'$,
where $\sigma'$ --- stress tensor of arounding medium,
we get
\begin{equation}\label{third}
{\rm div}\,(\sigma+\sigma')=0,
\end{equation}
that is local form of {\it third Newton law}.

Lets show curious origin of
second law in GR in terms of elasticity theory.
Assume, that space-time plate is characterized
by "phenomenological" multidimensional
elastic constant $E$ and $\sigma_P$. We'll
notate it shortly by
$S_{g}=S_{g}(E,\sigma_P)$, where $S_g$ --- is
action for gravitational field, or (up to a
dimensional constant) free elastic energy of bending (\ref{frenb}).
Action of matter we'll notate by $S_m$.
Variation of the full action $S=S_{g}(E,\sigma)+S_{m}$
over components of metric $g$ (or 4-dimensional strain tensor ${\cal D}$)
gives the sum of two tangent (to 4-dimensional space-time)
energy-momentum (stress) tensors
\begin{equation}\label{varS}
\delta S=\int(T^{(n)}(E,\sigma)+T^{(t)})
\cdot\delta gd\,{\rm vol}.
\end{equation}
The $T^{(n)}$ is borned by  normal straining of the plate, the $T^{(t)}$ ---
by a tangent ones.
Vanishing of (\ref{varS})
gives "generalized" Einstein equations, which
in terms of the stress tensors take the form:
\begin{equation}\label{non}
T^{(n)}(E,\sigma)+T^{(t)}=0.
\end{equation}
One can conclude, that Einstein theory
operates with  {\it nonstressed} state  of space-time.
In other words, physical meaning of standard Einstein equations
is expressed in intercompensation of tangent stresses,
owing to normal and tangent strainings.
From the view point of MET this is
not necessary equilibrium condition.
Equation (\ref{non}) has been obtained
under assuming, that components of $g$ (or ${\cal D}$)
are dynamical variables. However in elasticity theory such variables
are components of strain vector $\Xi$.
This alternative will be discussed below in Sec.(\ref{varpr}).
True equilibrium equations of the plate are:
\begin{equation}\label{eqein}
{\rm div}(T^{(n)}(E,\sigma)+T^{(t)})=0.
\end{equation}
Solution of (\ref{eqein}) describes equilibrium position
of the plate in embedding space and its stress distribution.
Lets go in (\ref{eqein}) to Einstein GR.
As it has been found in \cite{kok1} and mentioned
in Sec.(\ref{prev}) it can be made by setting $\sigma_{P}=1/2$.
In this condition $T^{(n)}(E,1/2)$ transforms (up to a constant) to
$-G/\ae=-({}^4{\cal R}{\rm ic}-(1/2)g{}^4{\cal R}_s)/\ae$ ---
pure geometrical
Einstein tensor (${}^4{\cal R}{\rm ic}_{\mu\nu}={}^4{\cal
R}^{\lambda}_{\mu\lambda\nu}$ --- Ricci tensor),
which because of Bianchi identities satisfies
${\rm div}\, G\equiv0$! Then from (\ref{eqein})
automatically follows ${\rm div} T^{(t)}\equiv0$
and we obtain well known statement
"equations of a motion are contained in
the field equations of GR".
Equation ${\rm div}\,T^{(t)}=0$,
when all kind of matter are included in $T^{(t)}$ expresses
second Newton law in covariant form.
So, we can conclude, that the fundamental principle
of classical mechanics can be associated with
the {\it special elastic properties
of space-time.}

\section{Problem of variational variables}\label{varpr}

Previous consideration has led to the question:
what relation will be between equations of the theory,
obtained from the same functional by
variating over different variables?
Let us consider it in details.

Let a system of field variables $\{q(x)\}$
considered in region $\Sigma$ of space-time
with metric $\theta$
is described by Lagrangian density $\Im(q,\dot q)$,
depending on field variables and its first derivatives
over coordinates, shortly abbreviated by $\dot q$.
Then a variation of action
$S=\int\limits_{\Sigma}\Im d\, {\rm vol}[\Sigma]$ over
variables $q$ will be given by the standard expression:
\begin{equation}\label{var1}
\delta S=\int\limits_{\Sigma}\delta_q\Im\delta q\, d\,{\rm vol}[\Sigma]+
\int\limits_{\partial\Sigma}\theta(\partial_{\dot q}\Im\, \delta q,
d\, {\rm vol}[\partial\Sigma])
\end{equation}
where  summation over $q$ and $\dot q$ is carried out if necessary,
$\delta_q$ ---
Euler-Lagrange operator, which gives field equation
$\delta_q\Im=0$.

Assume, that field variables depend on the set of potentials
$\{\xi\}$ through its derivatives $\xi'$ over coordinates:
$q=f(\xi')$.
Then variations of the field variables take the form:
$\delta q=
(\partial_{\xi'}f)\delta(\xi').
$
Substituting it  in (\ref{var1})
we get
\begin{equation}\label{var2}
\delta S=\int\limits_{\Sigma}\delta_q\Im
(\partial_{\xi'}f)\delta(\xi')
\,d\,{\rm vol}[\Sigma]+
\int\limits_{\partial\Sigma}\theta(\partial_{\dot q}\Im
(\partial_{\xi'}f)\delta(\xi'),\,
d\, {\rm vol}[\partial\Sigma])=
\end{equation}
\[
-\int\limits_{\Sigma}(\delta_q\Im
(\partial_{\xi'}f))'
\delta \xi\, d\,{\rm vol}[\Sigma]+
\int\limits_{\partial\Sigma}
\theta((\delta_q\Im)(\partial_{\xi'}f)\delta\xi,d\,{\rm
vol}[\partial\Sigma])+
\int\limits_{\partial\Sigma}\theta(\partial_{\dot q}\Im
(\partial_{\xi'}f)\delta(\xi'),
d\, {\rm vol}[\partial\Sigma])
\]
Comparing (\ref{var1}) and (\ref{var2}) we can conclude:
any solutions to field equations (\ref{var1}),
automatically are solutions to equations (\ref{var2}).
The same is valid for boundary conditions
and for the case of lagrangians with high derivatives.

Let us demonstrate it by the example with free
electromagnetic field. Its lagrangian (up to unessential in vaccum
constant)
has the
kind:
$\Im=F\cdot F=F_{\mu\nu}F^{\mu\nu}=2(\vec H^{2}-\vec E^{2})$.
Variation of action over  $\vec E, \vec H$
leads to the trivial equation of a motion:
$\vec E=0, \vec H=0$.
This solutions are simultaneously solutions to more general
system of free Maxwell equations,
which is  obtained by variation of action over 4-dimensional
vector-potential.

Returning to GR, we see, that {\it equations of
elasticity theory (\ref{perpeq})-(\ref{pareq})
together with the boundary conditions (\ref{bgen})
has  more generality
then Einstein Equations
from the viewpoint of variational procedure.}
So, what is called "exact solutions" is only subset of
solution of equilibrium equations, describing nonstressed
state of space-time.

\section{
Matter, its
energy-momentum tensor and signature of space-time.}\label{an}

Multidimensional elasticity approach throws new light
on meaning of matter, its energy momentum-tensor and energy density.

At first lets clarify geometric nature of an energy-momentum tensor.
From the view point of
second Newton law for point particle:
\begin{equation}\label{2new}
\vec a=\frac{\vec F}{m},
\end{equation}
force is vector, since acceleration $\vec a=\vec r_{tt}$
transforms as radii-vector. On the other hand, from
the view point of nature of fundamental  forces of
classical mechanics, force is 1-form $F$,
that follows from its relation with potential $U$:
$F(dx)=dU$. So, equation (\ref{2new}) should
be written in the following more exact form:
$$
\vec a=\frac{g^{-1}(\ ,F)}{m},
$$
where $g^{-1}$ --- contravariant metric.
Geometrical nature of hypersurface element $d\, {\rm
vol}[\partial\Sigma]$ is clear from
its local coordinate representation:
\begin{equation}\label{vol}
d\, {\rm vol}[\partial\Sigma]_{\mu}=\frac{\sqrt{{\rm det}[g]}}{(n-1)!}
\varepsilon_{\mu\mu_{1}\dots\mu_{n-1}}dx^{\mu_1}\wedge\dots
\wedge dx^{\mu_{n-1}}
\end{equation}
where $n$ --- dimension of $\Sigma$.
(\ref{vol}) shows, that $d\, {\rm vol}[\partial\Sigma]$ is 1-form too.
So, from the definition (\ref{stress}) of stress tensor $\sigma$
we get that last is
linear operator
$\hat\sigma$,
acting in space of 1-forms with tensor structure (1,1)\footnote{
(\ref{vol}) can be treated as dual conjugation of
hyphervolume $dx^{\mu_1}\wedge\dots\wedge dx^{\mu_{n-1}}$, so that
$\sigma$ become fully covariant tensor of valency $n$.
We'll use usual (dually conjugated) representation.
}.

Lets take for example energy-momentum tensor of isotropic perfect
fluid:
\begin{equation}\label{thd}
T=(p+\varepsilon)u\otimes u-pg,
\end{equation}
where $u$ --- 1-form of 4-velocity $(u=g(\vec u,\ ))$,
$p,\varepsilon$ --- scalars of pressure and energy density
in rest reference frame.
In this frame  energy-momentum affinnor $\hat T$  has the following kind:
\begin{equation}\label{mat}
\|\hat T\|={\rm diag}(\varepsilon,-p,-p,-p).
\end{equation}
In GR the statement
"there is a matter in some region $\Sigma$ of space-time"
is equivalent to the nonequality
$\varepsilon>0,\ x\in \Sigma$.
Value  $p$,
depend on  specific macroscopic properties of matter
and can be determined from equation of state.

As it has been discussed above, energy-momentum tensor of a matter
in terms of MET characterizes tangent stresses of space-time.
So, we go to the natural conclusion, that
{\it energy density
of a matter is nothing more no less then a pressure
of the plate medium in time-like direction,}
while  $p$ --- is common pressure in space-like
ones.
From the kind of (\ref{mat}) the following
fact can be deduced: {\it
signs  of pressures in time-like and space-like directions are
opposite}.
With the understanding, that pressure is positive,
if element of a volume
expands
under action of the pressure
forces
(matter tends to fly apart),
positive energy density will corresponds
to the negative pressure. Its negative signs
means, that substance of a plate undergoes contraction
along time-like direction.
In terms of 4-dimensional world we have bounded state of a matter
and anisotropic properties of space-time in time-like and
space-like direction. This pressures anisotropy obviously is own to
signature $(+1,-1,-1,-1)$ of local pseudoeuclidian metric, or
\begin{equation}\label{conseq}
\|\theta\|={\rm diag}(+1,-1,-1,-1)\ \longrightarrow
\|\hat T\|={\rm diag}(\varepsilon,-p,-p,-p).
\end{equation}
However, {\it physically it would be more properly
to determine the time-like direction as one,
along which the pressure of a plate has a negative sign.}
It means that arrow in (\ref{conseq}) should be reversed.
So, local hyperbolity of space-time, instead
of its inducing by metric of embedding space-time,
as it has been assumed in Sec.\ref{task} (expr. (\ref{sign})),
can be treated as
manifestation of signature anisotropy of tangent stresses
tensor in different 4-dimensional directions.

Note, that fully isotropic (in 4-dimensional sense)
liquid, which is called {\it false vacuume}, is widely used in
cosmology of early Universe \cite{cosm}.
It is described by equation of state
$p+\varepsilon=0$, and energy-momentum  tensor $\hat
T^{\mu}_{\nu}=-p\delta^{\mu}_{\nu}$,
which can be intrepreted as common
pascalian stress tensor in 4-dimensional euclidian space,
without time.

Note also, that such physical definition of time-like direction
nothing tells us about arrow of time, as it takes place
in special and general relativity too.

\section{Boundary conditions}\label{bd}

Majority works on field theory use freedom in definition
of lagrangian. Namely, two lagrangians, differing
by total divergence it is practice to consider equivalent
\cite{land2,bogol}.
Particulary, in variational procedure of GR,  total divergence terms, including second derivatives
of metric, are usually rejected \cite{land2}.
This is motivated
by vanishing of a metric variations  at a bound of
variation region. As a matter of a fact, this requirement is
perfectly unnecessary from the mathematical viewpoint
and has purely physical nature.
It is obviously, that adding
of divergence ${\rm div}{\cal V}$
to a
free energy density
${\cal F}$
will change boundary conditions:
\begin{equation}\label{div}
{\cal F}\to {\cal F}'={\cal F}+{\rm div}{\cal V}\ \mbox{\rm
means\ that}\
F'=F+F_{\rm surf}=\int\limits_{\Sigma}{\cal F}\,d\,{\rm vol}[\Sigma]+
\int\limits_{\partial\Sigma}\theta({\cal V},d\,{\rm
vol}[\partial\Sigma]).
\end{equation}
From the kind of last integral one can note,
that vector ${\cal V}$ has meaning of {\it vector surface energy
density}, which  physically can be related
with boundary effects (such as, for example, surface tension of liquid
drop). Stokes theorem lets write down this energy in the same
footing as volume energy.
Requirement of vanishing of variations at the bound means,
that certain choise of boundary condition is fixed.
In case of plate straining, conditions $\delta\Xi|_{\partial\Sigma}=0$
and $\delta(\partial\Xi)|_{\partial\Sigma}=0$ mean, that
edges of the plate are {\it pinned}. This choise
is typical for modern field theory.
Clearly, that this is not unique choise.
Vanishing of the first variation of  free energy, gives, besides of
equilibrium equations,
boundary conditions (\ref{bgen}), which can
be satisfied by different ways.
This conditions will determine the values
of integration constants and   a spectrum of eigen values
in Sturm-Liouville task. The most simple boundary conditions
at the plate edges become in cases of {\it pinned,
simply supported or free bounds} \cite{land1}.
The case of unbounded
in one or several dimensions plate
is also sufficiently simple.

However, such arbitrary choice of boundary conditions
is speculative, or at best, model.
{\it The problem of boundary conditions must be
disolved by an experimental way, if we are talking about physical theory.}
So, the role of dynamical equations should be expanded:
its solutions after comparing with
observational dates can be used not only for
describing of dynamics of gravitational and other fields,
but also for investigation of boundary conditions
of our world.
In paper \cite{kok2} the
case of rectangle plate with different
combinations of boundary conditions has been considered
in cosmological aspects.
It has been shown,
that measurable
4-dimensional geometric invariants are
sensible (in some extent) to a choise  of boundary conditions.

It is remarkable, that gauge fixing in gauge theories
concerns to choise of boundary conditions too.
It can be easily seen in the simplest case of $U(1)$-symmetry
of electrodynamics.
Action of interaction $S_{\rm int}$ in electrodynamics has the form
\begin{equation}\label{el}
S_{\rm int}=\int\limits_{\Sigma} A(j)\, d\, {\rm vol}[\Sigma],
\end{equation}
where $A$ and $j$ --- 1-form of potential and vector of electric
current correspondingly.
Let make gauge transformation $A\to A+\nabla\chi$,
where $\chi$ --- arbitrary smooth scalar function.
Then (\ref{el}) transforms to
$$
S'_{\rm int}=S_{int}+\int\limits_{\Sigma}d\chi(j)\, d\,{\rm vol}[\Sigma]=
S_{\rm int}+\int\limits_{\Sigma}{\rm div}\,(\chi
j)\, d\,{\rm vol}[\Sigma] -\int\limits_{\Sigma}\chi{\rm div}\,
j\,d\,{\rm vol}[\Sigma].
$$
Vanishing of the last integral gives conservation law of
charge: ${\rm div}j=0$, while second integral can be transformed
into surface term:
$$
\oint_{\partial\Sigma}\chi\theta(j,d\,{\rm vol}(\partial[\Sigma])).
$$
Its vanishing means special choise of boundary conditions
for vector $\chi j$, playing role of surface energy density.
So, gauge invariance of action
(and conservation of electric charge) is closely related
with boundary conditions for fields and its sources.

\section{Conclusion}

Proposed approach gives unified base for description of gravity,
field and matter. As standard elasticity theory, the approach has
apparently "phenomenological" kind, since
1) postulate about Hookes law validity is accepted;
2) phenomenological Lame coefficients should be measured
experimentally.
Quotation-marks mean, that this multidimensional
phenomenologicity is manifested as
fundamental laws of 4-dimensional world
(see Eq.(\ref{perpeq}), (\ref{pareq})). From the other hand,
any classical field theory (for example Einstein GR)
can be treated as phenomenological, since 1)
it postulates lagrangian (in view (MET) --- free elastic energy
functional),
whose simple kind has the similar to Hookes law nature;
2) it postulates set of fundamental constants,
plaing role of Lame coefficient.
So, theories of gravity
with nonlinear over curvature terms, can
be regarded as MET with generalized
Hookes law.

There is many deep analogies of MET with
modern fundamental theories, mentioned in Introduction.
Detail discussion we defer to the future.

Lets shortly discuss problems of motion and quantization.
Motion of any finite body in 4-dimensional world
can be represented by its world tube. From the view
point of MET this last should be considered as {\it thin elastic
bar}, generally speaking, {\it bent} and {\it twisted.}
Bending is determined by acceleration of mass center,
twisting --- by rotation of the body.
We can accept the following natural hypothesis:
{\it standard mechanical action $S=\int{\cal L}\,dt=\int(K-U)\,dt$,
where $K$ and $U$ --- kynetic and potential energy correspondingly
has the sense of free energy of bending and twisting of the elastic
bar.}
Preliminary investigation supports this idea.
Another possibility, which indirectly follows from the previous
consideration,
is related to the theory of {\it nemathic medium.}
Anisotropy of energy-momentum tensor in time-like and space-like
directions can be regarded as manifestation of
{\it nemathic properties of space-time
plate.} This point of view demands multidimensional generalization
of standard nemathic theory (see \cite{hehl}).

Problems of time arrow and of observer's motion along time-like
world line remain opened (as in
GR and majority of its generalization). Presumably, it can'nt
be resolved in classical physics (see \cite{pav}).

It would be naturally to believe, that
quantum phenomena should be related to
a small finite size of the plate thicknesses. One can consider
the set of thicknesses $\{h_m\}$ as a set of scales,
which {\it determine elementary stable stressed states of different
sort, observing as elementary particles.} This idea is closely
related with results of work \cite{dmitr}.

\vspace{2cm}
{\bf Acknowledgement}

I am most grateful to V.A.Korotky and participants of
seminar "Geometry and physics" for useful discussions.

\appendix

\section{Existing of vector field  $a$}.

Lets prove existence of vector field $a$,
satisfying the following conditions:

1) $a$ depends on coordinate only through scalar field $\phi$:
$a=a(\phi)$;

2) $a(0)=0$;

3) $\theta(a_{\phi},a_{\phi})=const$, where $a_{\phi}\equiv
da/d\phi$,  $\theta$ --- Minkowski metric.

4) ${\rm div}\,a=0$.

Since all conditions (1)-(4) are coordinate independent,
we can work in any coordinate system.
Lets begin from
point (4).
After changing coordinates\footnote{Without loss
of generality we consider $d\phi$ as time-like
1-form. In case of space-like and isotropic $d\phi$
proof can be carried out by the same way.}:
$x'^0=\phi(x),\ x'^{1,2,3}=x^{1,2,3}$, condition
(4) will take the form (we ommit "$'$"):
$$
{\rm div}a=d\phi(a_{\phi})=a^0_{\phi}=0.
$$
So, (4) will be satisfied, when $a=(c_0,\vec a)$ in the
special coordinate system ($\vec a=(a^1,a^2,a^3),\ a^i$ ---
any differentiable functions of $\phi$, $c_0=const$).
Put $c_0=0$, then (3) means $\vec
a_{\phi}^2=-(a^1)^2-(a^2)^2-(a^3)^2=-R^2=const$.
Going to spherical angles $\vartheta=\vartheta(\phi)$,
$\varphi=\varphi(\phi)$, we get:
$$
\left\{
\begin{array}{lcr}
a^1_{\phi}=R\sin\theta\cos\varphi;\\
a^2_{\phi}=R\sin\theta\sin\varphi;\\
a^3_{\phi}=R\cos\theta,
\end{array}
\right.
\mbox{\rm and\ integrating}\
\left\{
\begin{array}{lcrcl}
a^1&=&R\int\sin\theta\cos\varphi\,d\phi+c_1&=&RF_1(\phi)+c_1;\\
a^2&=&R\int\sin\theta\sin\varphi\,d\phi+c_2&=&RF_2(\phi)+c_2;\\
a^3&=&R\int\cos\theta\,d\phi+c_3&=&RF_3(\phi)+c_3.
\end{array}
\right.
$$
Putting $c_i=-RF_i(0)$ we satisfy remained condition (2).

\end{document}